\documentclass{PoS}
\usepackage{amsmath}
\usepackage[numbers,sort&compress]{natbib}

\title{Confronting the Inert Doublet Model with results from Run 1 of the LHC}

\ShortTitle{Confronting the Inert Doublet Model with results from Run 1 of the LHC}

\author{\speaker{D. Sengupta}
         \thanks{In collaboration with : G. Belanger, B. Dumont, A. Goudelis, B. Herrmann and S.Kraml}\\
         Laboratoire de Physique Subatomique et de Cosmologie, Universit\'e Grenoble-Alpes, 
CNRS/IN2P3, 53 Avenue des Martyrs, F-38026 Grenoble, France \\
        E-mail: \email{dipan.sengupta@lpsc.in2p3.fr}}

\abstract{ The Inert Doublet Model (IDM) is a simple extension of the Standard Model (SM)
that aims to address the naturalness problem, electroweak baryogenesis and accommodate
a viable dark matter (DM) candidate, along with a rich phenomenlogy in terms of collider 
signatures. In this note, we address the constraints on the IDM 
from dilepton searches performed at LHC run-1.}

\FullConference{18th International Conference From the Planck Scale to the Electroweak Scale \\
                 25-29 May 2015\\
                 Ioannina, Greece \\
   }

\begin{document}

\section{Introduction}
The IDM, a simple extension of the two Higgs doublet model (2HDM) by 
an additional $\mathbb Z_{2}$ symmetry, has been 
studied for almost 4 decades in various contexts. Initially introduced as 
a model for electroweak symmetry breaking \cite{IDMfirst}, the model has subsequently been 
found useful to address a whole array of issues, including improved naturalness \cite{IDMnaturalness},
baryogenesis\cite{Ma:2006fn}, generation of neutrino masses via see-saw mechanism \cite{Ma:2006km}, and 
importantly  a viable dark matter candidate predicting the correct relic 
density of the universe \cite{IDMarchetype,IDMCW,IDMnewviable}. The IDM provides a rich phenomenology that has been 
previously studied both in the context of LHC signatures \cite{IDMLHChinvfirst,IDMdileptons1,IDMhgaga1}, as well as astrophysical
data on direct and indirect detection of dark matter \cite{IDMgammalines,IDMnu1,IDMnu2,IDMmultileptons}. Furthermore, theoretical 
constraints like vacuum stability, perturbativity, as well as constraints originating 
from electroweak precision data and dark matter
 have also been studied in some detail \cite{Khan:2015ipa,Swiezewska:2015paa,IDMpostHiggs}. The discovery of the 125 GeV Higgs boson \cite{ATLASDiscovery,CMSDiscovery} has constrained the 
the IDM parameter space significantly.
However,  a study of constraints coming from the direct searches performed during
 LHC run-1  has not been conducted. 
In this work \cite{Belanger:2015kga}, we analyse the constraints on the IDM parameter space from the dilepton +
missing energy ($E_{T}^{miss}$) searches performed at LHC run-1. We recast two publicly 
available ATLAS analyses for the Electroweakino search \cite{ATLASsusy} and invisible Higgs \cite{ATLASinvh}
in the MadAnalysis5 \cite{Conte:2014zja,Conte:2012fm} framework,
 now available on the public analysis database \cite{Dumont:2014tja}, 
 and reinterpret  them to constrain the  IDM parameter space. 
 
 \section{IDM parameter space}

 The IDM is an extension of the SM by the addition of a second SU(2) doublet scalar $\Phi$, 
 being  odd under the additional $\mathbb {Z}_{2}$ symmetry.  The two doublets can be 
 written as, 
 \begin{equation}
	H = \left( \begin{array}{c} G^+ \\ \frac{1}{\sqrt{2}}\left(v + h + i G^0\right) \end{array} \right),
	\
	\Phi = \left( \begin{array}{c} H^+\\ \frac{1}{\sqrt{2}}\left(H^0 + i A^0\right) \end{array} \right).
\end{equation}

The SM like Higgs boson is represented by h, while $ H^{0}$ and $A^{0}$ denote the additional 
CP even neutral and CP odd neutral scalars respectively. Additionally it contains a pair charged 
scalars ($H^{\pm}$).  $G^{0}$ and $G^{+}$ represent the  neutral and the charged goldstone 
bosons respectively.  With these field definitions, the scalar part of the potential has the form, 
   \begin{align}
	V_0 & = \mu_1^2 |H|^2  + \mu_2^2|\Phi|^2 + \lambda_1 |H|^4+ \lambda_2 |\Phi|^4 \\ \nonumber
		& + \lambda_3 |H|^2| \Phi|^2 + \lambda_4 |H^\dagger\Phi|^2 + \frac{\lambda_5}{2} \Bigl[ (H^\dagger\Phi)^2 + \mathrm{h.c.} \Bigr].
\label{Eq:TreePotential}
\end{align}

The physical masses for the particles, given in terms of the Largrangian parameters read, 
\begin{align}
	m_{h}^2 &= \mu_1^2 + 3 \lambda_1 v^2, \\ 	
	m_{H^0}^2 &= \mu_2^2 + \lambda_L v^2, \label{Eq:mH0tree} \\
	m_{A^0}^2 &= \mu_2^2 + \lambda_S v^2, \\
	m_{H^{\pm}}^2 &= \mu_2^2 + \frac{1}{2} \lambda_3 v^2, 
\end{align}
with the couplings $\lambda_{L,S}$  defined as,
\begin{eqnarray*}
	\lambda_{L,S} &= \frac{1}{2} \left( \lambda_3 + \lambda_4 \pm \lambda_5 \right).
\end{eqnarray*} 

The $\rm \mathbb{Z}_{2}$, symmetry ensures that the doublets H and $\rm \Phi$
do not mix. A further consequence is that the lightest 
$\rm \mathbb{Z}_{2}$ odd particle (LOP) ($H^{0}$ or $A^{0}$), 
 can be a candidate for a dark matter particle.

The following constraints are imposed on the IDM parameter space in 
this work, 

\begin{enumerate}
\item The first set of constraints arises from theoretical requirements like 
stability of the electroweak vacuum, validity of perturbativity and unitarity
of the model \cite{Khan:2015ipa,Swiezewska:2015paa}. These three requirements are assumed to be valid up to 10 TeV. 

\item The next set of constraints imposed are on the contributions of the IDM
to the oblique parameters $S,T,U$. We use 3 $\sigma$ ranges as described in \cite{STUparams}.

\item Furthermore we impose LEP constraints from the neutralino and chargino searches \cite{IDMLEPII}. 
We require (assuming $m_{H^0} < m_{A^0}$) $m_{A^0} \gtrsim 100$ GeV and $m_{H^\pm} \gtrsim m_W$.

\item For $m_{H^0} \leq m_h/2$, (without the loss of generality) we must ensure that the constraints arising from the 
decay modes of the SM Higgs are in accordance to the latest experimental values. Assuming that 
the Higgs has SM-like couplings, the invisible branching ratio of the Higgs to two $H^0$
 particles is constrained to ${\rm BR}(h\rightarrow {\rm inv.}) < 0.12$ at $95\%$~confidence level
 (CL)~\cite{StatushCouplingsrun1}
 (see also \cite{StatusHiggsInv,HiggsAtLast,HiggsSignalStrengths2013,HiggsSignalsStudy}).   
In this work, we work in the limit of $\lambda_{L} \rightarrow 0$, where constraints
 arising from invisible Higgs decays vanish. 

\end{enumerate}

Finally we assess the dark matter constraints on the IDM.
 If the IDM is considered as a viable dark matter model, it can satisfy the
observed DM relic density \cite{Planck2013} in three regions: the low mass regime, $m_{H^0} < m_W$, where $\lambda_L$,
 the coupling relating the two $H^0$  particles to a Higgs boson, and the mass difference 
  $m_h/2 - m_{H^0}$ plays a crucial role \footnote{the exact difference between $m_{H^0}$
 and $m_W$ also plays a role when the former is larger than $\sim 70$ GeV}; the intermediate mass
region $m_W < m_{H^0} \lesssim 115$ GeV, where the relic density is governed by $m_{H^0}$ and $\lambda_L$;
and finally the high mass region where all parameters of the scalar potential except
 $\lambda_2$ drastically affect the DM relic abundance. In this work the high mass region is not of interest. 
In the limit of $\lambda_{L} \rightarrow 0$, constraints
 arising from invisible Higgs decays vanish \footnote{The exact constraint is $\lambda_{L} \le 6\times 10^{-3}$, 
above which the entire low mass region is ruled out by $h\to inv$.}.
 Furthermore, XENON100 \cite{XENON100225days} has already eliminated the entire low-to-intermediate mass regime, 
where the IDM can satisfy the observed relic density according to the freeze out mechanism.
Thus with the exception of a highly fine tuned Higgs funnel region, the $\lambda_L\to 0$ regime leads to a DM 
overabundance \cite{IDMpostHiggs}. However one can always argue that the IDM can be treated as a model with interesting 
collider phenomenology without strictly imposing a DM constraint. In view of this we assess the constraints 
on the IDM from the LHC dilepton searches.

\section{LHC constraints on the IDM }

To assess the constraints on the IDM we considered the dilepton and the trilepton channels 
as the primary targets because of lower backgrounds and a cleaner signature. However 
the trilepton channel suffers significantly due to lower cross sections. Hence we 
consider the dilepton channel in this work. 
 
 The following four channels yielding a dilepton + $E_{T}^{miss}$ final state were considered in this 
work, 
\begin{align} \label{eq:AH0}
q \bar{q} & \rightarrow Z \rightarrow A^0 H^0 \rightarrow Z^{(*)} H^0 H^0 \rightarrow \ell^+ \ell^- H^0 H^0 , \\ \label{eq:HpHm}
q \bar{q} & \rightarrow Z \rightarrow H^{\pm} H^{\mp} \rightarrow W^{\pm(*)} H^0 W^{\mp(*)} H^0 \\ \nonumber 
          & \rightarrow \nu \ell^+ H^0 \nu \ell^- H^0 , \\ \label{eq:Zh}
q \bar{q} & \rightarrow Z \rightarrow Z h^{(*)} \rightarrow \ell^+ \ell^- H^0 H^0 , \\ \label{eq:fourvertex}
q \bar{q} & \rightarrow Z \rightarrow Z H^0 H^0 \rightarrow \ell^+ \ell^- H^0 H^0 .
\end{align}  

Although no dedicated search for the IDM has been performed at the LHC, ATLAS and CMS collaborations
have performed studies with the same final state in the context of SUSY and invisible Higgs search \cite{ATLASsusy}.
These searches can be therefore be reinterpreted to constrain the IDM parameter space. To this end, we pick up 
two ATLAS searches, and recast them in the framework of {\sc MadAnalysis~5} \cite{Conte:2012fm,Conte:2014zja}. 
The first of these is the ATLAS opposite side dilepton + $E_{T}^{miss}$
 performed for the search of electroweakinos and sleptons \cite{ATLASsusy}.The $\ell^+\ell^- + E_{T}^{\rm miss}$ 
signature can arise from 
chargino-pair production followed by $\tilde\chi^\pm\to W^{\pm(*)} \tilde\chi^0_1$ or
 $\tilde\chi^\pm\to \ell^\pm \tilde\nu / \tilde\nu \ell^\pm$ decays, or 
slepton-pair production followed by $\tilde\ell^\pm\to \ell^\pm \tilde\chi^0_1$ decays. 
For this analyses \cite{ATLASsusy} the signal regions (SR) rely on purely leptonic final state.
 The analyses requires a veto on the Z boson to reduce the ZZ background, with 
$|m_{\ell\ell}-m_Z| > 10$~GeV. The $\ell^+\ell^- + E_{T}^{\rm miss}$ signature corresponding 
to the simplified model $\tilde\chi^+\tilde\chi^- \to W^+(\to \ell^+\nu)\tilde\chi^0_1 W^-(\to \ell^-\nu)\tilde\chi^0_1$
considered in the analyses  can be directly mapped the IDM channel described in Eq. ~\ref{eq:HpHm}.

The second ATLAS search of interest is the search for invisible decays of a Higgs boson produced in association with 
a $Z$ boson~\cite{ATLASinvh}. This search is complementary to the electroweakino search in accepting 
the reconstructed Z boson from the $\ell^{+}\ell^{-}$ pair, the specific requirement being $|m_{\ell\ell}-m_Z| < 15$~GeV. 
This process is a direct analogue of Eq. \ref{eq:Zh} and \ref{eq:fourvertex}.  Additionally, a dark matter search 
with $\ell^+\ell^- + E_{T}^{\rm miss}$ \cite{ATLASdm} final state was also performed. However the analyses requires
a large missing energy criteria which wipes out the entire signal.

As noted earlier, these two ATLAS analyses are recasted using the 
{\sc MadAnalysis~5}~\cite{Conte:2012fm,Conte:2014zja} framework. 
While the SUSY search~\cite{ATLASsusy} was already 
available in the Public Analysis Database~\cite{Dumont:2014tja} as the recast code~\cite{ma5susy}, 
the invisible Higgs search~\cite{ATLASinvh} was implemented and validated for this work 
and is now available at~\cite{ma5Higgs}. For the recast, signal samples were generated 
with Madgraph5 \cite{Alwall:2011uj} using the model implemented using Feynrules 
 available in \cite{IDMpostHiggs}, with the widths calculated by Calchep \cite{Pukhov:2004ca,Belyaev:2012qa}.
The parton-level events were passed on to {\sc Pythia}~6.4~\cite{Sjostrand:2006za} for  
showering and hadronization.  The `MA5tune' version of {\sc Delphes}~3~\cite{delphes3} 
(see Section~2.2 of \cite{Dumont:2014tja}) was used for detector simulation.  
The cutflows were obtained with the recast codes described in~\cite{ma5susy,ma5Higgs}. 
To obtain a  statistical interpretation, we used the module {\tt exclusion\_CLs.py}~\cite{Dumont:2014tja}. 
With the given  number of signal, observed and expected 
background events, along with the background uncertainty,  {\tt exclusion\_CLs.py} determines 
the most sensitive SR, the exclusion confidence level using the $\mathrm{CL}_s$ prescription, 
and the nominal cross section $\sigma_{95}$ that is excluded at 95\%~CL.\footnote{Note that we do not 
simulate the backgrounds but take the background numbers and uncertainties  directly from the 
experimental publications \cite{ATLASsusy} and \cite{ATLASinvh}.}

 With the imposition of $\lambda_L = 0$, 
and  noting that $\lambda_2$ is irrelevant for all observables at tree-level, we are left with
 $m_{A^0}$ and $m_{H^0}$  to scan over. For $m_{H^\pm}$, we choose two representative values:
 $m_{H^\pm} = 85$~GeV, which is the lower allowed limit by LEP, 
and $m_{H^\pm} = 150$~GeV, which is significantly higher but still safely within the bounds imposed by the $T$ parameter,
 which limits the mass splitting between the inert scalar states (see also the analysis in \cite{IDMpostHiggs}).
 
\begin{figure}
\includegraphics[width=9cm]{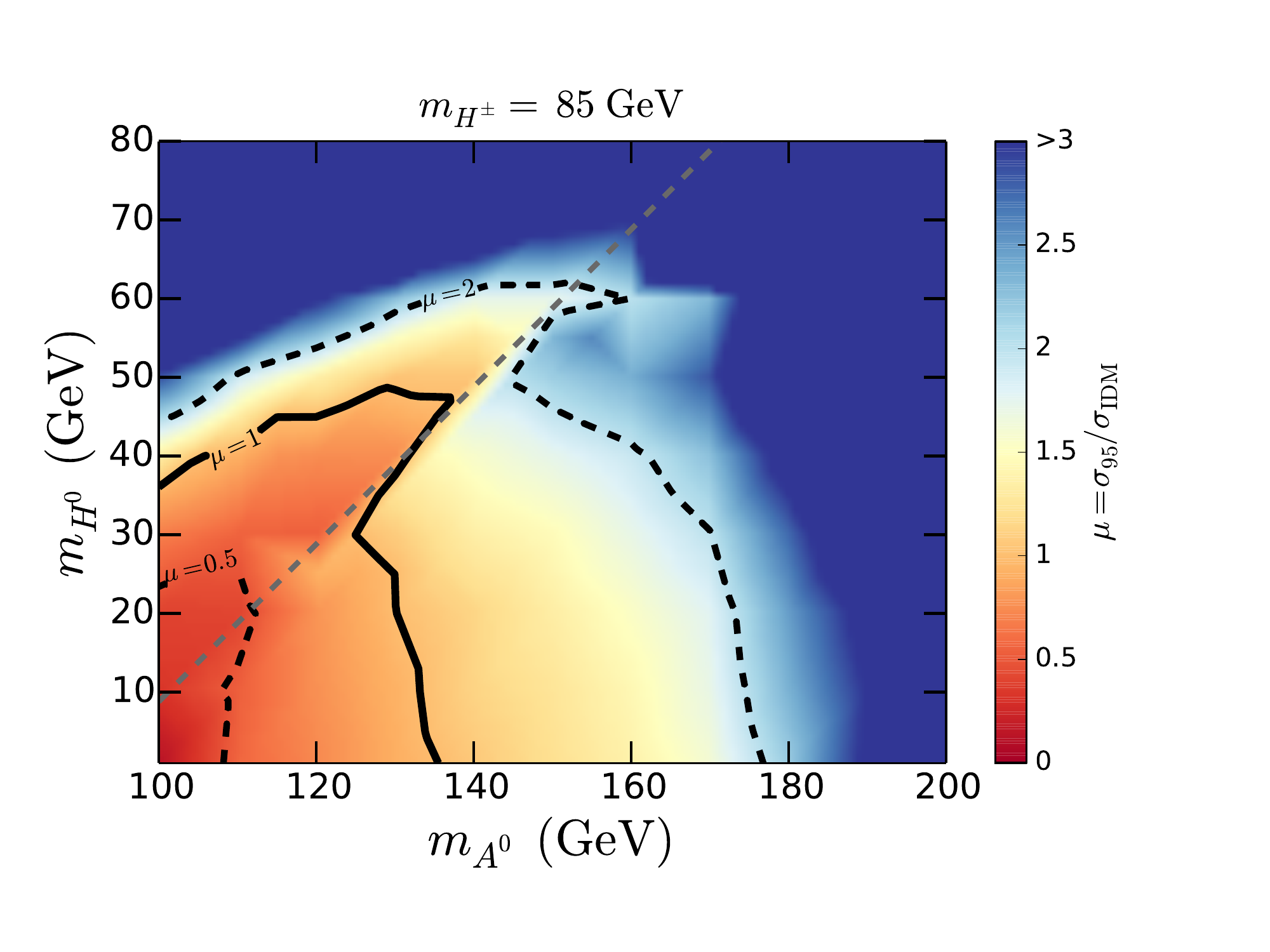}\vspace*{-10mm}
\includegraphics[width=9cm]{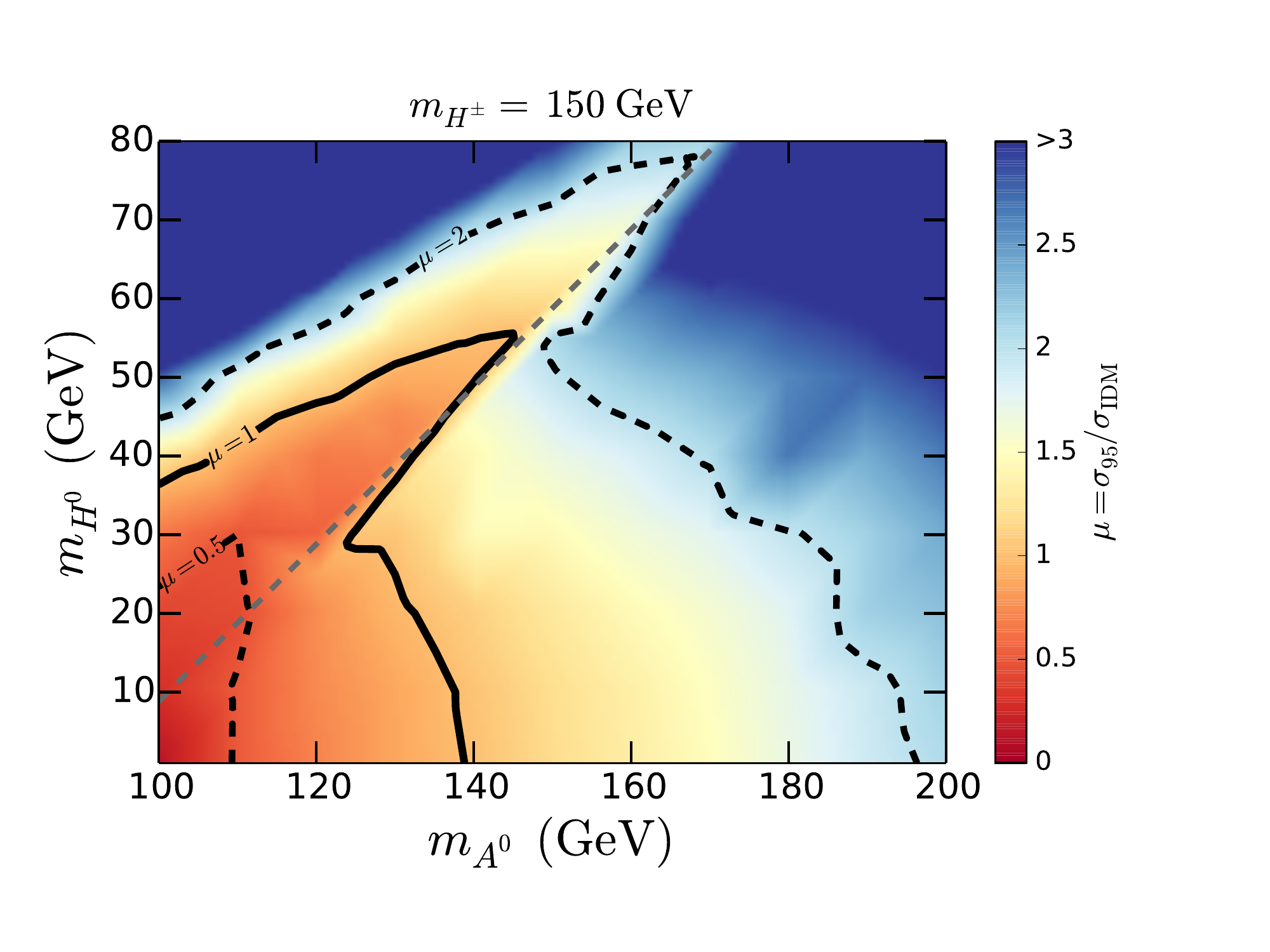}
\caption{The ratio $\mu \equiv \sigma_{95}/\sigma_{\rm IDM}$ in the $(m_{A^0}, m_{H^0})$ plane for two representative values of the charged inert scalar mass, $m_{H^\pm} = 85$ GeV (upper panel) and $m_{H^\pm} = 150$ GeV (lower panel). The solid black lines are the 95\%~CL exclusion contours, $\mu = 1$. The dashed black lines are given for illustration and correspond to the $\mu = 0.5$ and $\mu = 2$ contours. The grey dashed lines indicate $m_{A^0}-m_{H^0}=m_Z$. }
\label{fig:exclusionplots}
\end{figure}

The result for the recasted analyses is presented in Fig.~\ref{fig:exclusionplots}, where we show 
$\mu \equiv \sigma_{95}/\sigma_{\rm IDM}$
in the form of temperature plots in the $(m_{A^0}, m_{H^0})$ plane for the two chosen values of $m_{H^\pm}$. 
In these plots $\sigma_{\rm IDM}$ is the cross section predicted by the model while $\sigma_{95}$ is the cross section 
excluded at 95\%~CL. Thus regions with $\mu \le 1$ are excluded at 95\%~CL.

We observe that Run~1 ATLAS dilepton searches exclude inert scalar masses up to about $35$~GeV for pseudoscalar masses around $100$~GeV
at 95 $\%$ CL.  The limits become stronger for larger $m_{A^0}$, with $m_{H^0}$ $\approx 45$ (55)~GeV ruled out for $m_{A^0} \approx 140$ (145)~GeV and $m_{H^\pm} = 85$ (150)~GeV. 
For massless $H^0$, $A^0$ masses up to about 135--140~GeV are excluded  
( $m_{H^0}$ and $m_{A^0}$ are generally interchangeable here).
It can be observed that the constraints are slightly stronger for heavier $m_{H^\pm} $. 
This is because of sub-leading contributions from Eq. \ref{eq:HpHm} and from $q \bar{q} \to W^\pm \to AH^{\pm} \to Z^{(*)}H W^{\pm(*)}H$ with a missed lepton. Although the cross section is much larger for  $m_{H^\pm} = 85$~GeV as compared to $m_{H^\pm} = 150$~GeV, the resulting leptons are much softer and generally do not pass the signal requirements.

It can also be observed the limits on $m_{H^0}$ become stronger for larger $A^0$ masses. 
 This is  because the leptons originating from the $A^0\to Z^{(*)}H^0$ decay are harder with more available phase space
  and thus pass the signal selection cuts more easily.
 On the other hand for smaller mass splittings between $H^0$ and $A^0$,
 the produced dileptons are much softer. 

On the other hand, when  $m_{A^0}-m_{H^0}\ge m_{Z}$, the $Z$ boson is on-shell, and hence the $Z$ veto in the SUSY analysis eliminates most of the signal. In this region the invisible Higgs analsyses, $Zh\to \ell^+\ell^- + E_T^{\rm miss}$ provides the stronger limit. 
 Although we use $m_{H^\pm}=150$~GeV in this work,
  the 
mass of the charged scalar has a negligible effect in the final exclusion. 
 It can be noticed that there is a
 small overlap region at low $H^0$ mass, $m_{H^0}\lesssim 25$~GeV,
  where the SUSY search constrain a small region
despite the $Z$ boson from the $A^0$ decay being on-shell.
  This comes from the significantly large tail of the $m_{\ell\ell}$ distribution extending
 below $m_{\ell\ell}=m_Z-10$~GeV, where some signal events are picked up in
 the SUSY SR $WWb$ (this tail was already noticed in \cite{IDMdileptons1}).

Finally a discussion on the prospects of the 13 TeV Run is worthwhile. With the minimal assumption that one can na\"ively rescale  signal and background numbers (see, \textit{e.g.}, \cite{Bharucha:2013epa}),
 (assuming that the acceptance$\times$efficiency values remain the same) indicates that at 13~TeV with an integrated luminosity of 100~fb$^{-1}$, the 95\%~CL reach can be extended up to $\mu\approx1.2$ (1.6) above (below)
 the dashed grey line in Figure~\ref{fig:exclusionplots}.
 This reach improves to $\mu\approx2.1$ (2.7) with a luminosity of  300~fb$^{-1}$,
 hence covering a major part of 
the $(m_{A^0}, m_{H^0})$ plane shown in Figure~\ref{fig:exclusionplots}. This reach, however
depends on a number of factors including detector performance affecting the signal and background 
estimation, the actual change in kinematics at an increased energy. Furthermore, we advocate that 
a dedicated search should be devised to investigate this model, with an analysis targeted 
specifically for $pp\rightarrow A^0H^0\rightarrow Z^{(*)} H^0H^0$ (or alternatively $pp\rightarrow \tilde\chi_2^0\tilde\chi_1^0\rightarrow Z^{(*)}\tilde\chi_1^0\tilde\chi_1^0$). Variables like angular separation between leptons
can be helpful in this respect. 

\section{Conclusion}
In this work we addressed the constraints on the Inert Doublet Model 
coming from dilepton + $E_{T}^{miss}$ searches at LHC.
We  conclude that the Run 1 of LHC provide significant
constraints on the IDM parameter space, extending them beyond 
the LEP, and complementary to cosmological and astrophysical 
constraints. The 95\%~CL limits  derived from the dilepton $+\; E_T^{\rm miss}$ Higgs and SUSY analyses
 exclude inert scalar masses of up to about 55~GeV in the best cases.
It should be possible to push these constraints beyond $m_{h}/2\approx 62.5$~GeV at Run~2 of the LHC, thus enabling
  to probe the Higgs funnel region of the IDM, a regime. 
However a dedicated experimental analysis for the IDM at 13~TeV would be highly desirable, 
as it would enable to constrain the parameter space beyond the recasted analysis with optimized 
cuts.

\section{Acknowledgements}
This work was supported in part by the ANR project {\sc DMAstroLHC}, ANR-12-BS05-0006, the 
``Investissements d'avenir, Labex ENIGMASS'', and by the IBS under Project Code IBS-R018-D1.

\bibliography{skeleton}{}
\bibliographystyle{utphys.bst}
\end{document}